\documentclass{JHEP3}
\usepackage{graphicx,amsmath,amssymb}
\usepackage{epsfig,multicol}
\usepackage{epsf}
\usepackage{epstopdf}
\input{epsf.sty}

\usepackage{graphicx}
\usepackage{amssymb}
\usepackage{amsmath}

\def\p{\partial}

\def\half{{1\over 2}}
\def\({\left(}
\def\){\right)}
\def\[{\left[}
\def\]{\right]}

\def\e{\begin{equation}}
\def\q{\end{equation}}
\def\m{\begin{eqnarray}}
\def\n{\end{eqnarray}}

\def\E{{\bf e}}

\title{A geometric description of the non-Gaussianity generated at the end of multi-field inflation}
\author{Qing-Guo Huang \footnote{huangqg@kias.re.kr}
\\\small{\em School of Physics, Korea Institute for Advanced Study,
207-43, Cheongryangri-Dong, Dongdaemun-Gu, Seoul 130-722, Korea }
\\\small{\em Kavli Institute for Theoretical Physics China,
ITP-CAS, Beijing, P.R. China}}

\abstract{
In this paper we mainly focus on the curvature perturbation generated at the end of multi-field inflation, such as the multi-brid inflation. Since the curvature perturbation is produced on the super-horizon scale, the bispectrum and trispectrum have a local shape. The size of bispectrum is measured by $f_{NL}$ and the trispectrum is characterized by two parameters $\tau_{NL}$ and $g_{NL}$. For simplicity, the trajectory of inflaton is assumed to be a straight line in the field space and then the entropic perturbations do not contribute to the curvature perturbation during inflation. As long as the background inflaton path is not orthogonal to the hyper-surface for inflation to end, the entropic perturbation can make a contribution to the curvature perturbation at the end of inflation and a large local-type non-Gaussiantiy is expected. An interesting thing is that the non-Gaussianity parameters are completely determined by the geometric properties of the hyper-surface of the end of inflation. For example, $f_{NL}$ is proportional to the curvature of the curve on this hyper-surface along the adiabatic direction and $g_{NL}$ is related to the change of the curvature radius per unit arc-length of this curve. Both $f_{NL}$ and $g_{NL}$ can be positive or negative respectively, but $\tau_{NL}$ must be positive and not less than $({6\over 5}f_{NL})^2$.
}

\preprint{CAS-KITPC/ITP-105}


\begin{document}

\section{Introduction}

In the single-field slow-roll inflation the different Fourier components of the curvature perturbation are roughly uncorrelated and their distribution is almost Gaussian \cite{Maldacena:2002vr}. However in the fundamental theory going beyond the standard model, such as string theory, a large number of scalar fields are expected and the distribution of the fluctuations can significantly deviate from a  Gaussian distribution. A well-understood ansatz of non-Gaussianity has a local shape \cite{Komatsu:2000vy}. Working in the framework of Fourier transformation of the curvature perturbation $\zeta$, the primordial power spectrum of the curvature perturbation ${\cal P}_\zeta$ is defined by 
\e 
\langle\zeta({\bf k_1})\zeta({\bf
k_2})\rangle=(2\pi)^3 {\cal P}_{\zeta}(k_1)\delta^3({\bf k_1}+{\bf
k_2}), 
\q  
and the primordial bispectrum and trispectrum are respectively 
\m
\langle\zeta({\bf k_1})\zeta({\bf k_2})\zeta({\bf
k_3})\rangle&=&(2\pi)^3 B_\zeta(k_1,k_2,k_3)\delta^3({\bf k_1}+{\bf k_2}+{\bf k_3}), \\
\langle\zeta({\bf k_1})\zeta({\bf k_2})\zeta({\bf k_3})\zeta({\bf
k_4})\rangle&=&(2\pi)^3 T_\zeta(k_1,k_2,k_3,k_4)\delta^3({\bf
k_1}+{\bf k_2}+{\bf k_3}+{\bf k_4}). 
\n 
The bispectrum and
trispectrum are respectively related to the power spectrum by 
\m
B_\zeta(k_1,k_2,k_3)&=&{6\over 5}
f_{NL}[{\cal P}_\zeta(k_1){\cal P}_\zeta(k_2)+2\ \hbox{perms}], \label{bi} \\
T_\zeta(k_1,k_2,k_3,k_4)&=&\tau_{NL}[{\cal P}_\zeta(k_{13}){\cal
P}_\zeta(k_3){\cal P}_\zeta(k_4)+11\ \hbox{perms}] \nonumber \\
&+&{54\over 25}g_{NL}[{\cal P}_\zeta(k_2){\cal P}_\zeta(k_3){\cal
P}_\zeta(k_4)+3\ \hbox{perms}], \label{tri} 
\n 
where $f_{NL}$, $\tau_{NL}$ and $g_{NL}$ are the non-Gaussianity parameters which measure the size of the deviation from the Gaussian distribution.

Using the $\delta N$ formalism \cite{Starobinsky:1986fxa}, the non-Gaussianity parameters can be written by
\m
f_{NL}&=&{5\over 6}{N_{,ij}N_{,i}N_{,j}\over (N_{,l}N_{,l})^2}, \label{nfnl}\\
\tau_{NL}&=&{N_{,ij}N_{,ik}N_{,j}N_{,k}\over (N_{,l}N_{,l})^3}, \label{ntaunl} \\
g_{NL}&=&{25\over 54} {N_{,ijk}N_{,i}N_{,j}N_{,k}\over (N_{,l}N_{,l})^3}, \label{ngnl}
\n
where $N$ is the number of e-folds before the end of inflation, and 
\e
N_{,i}={\p N\over \p\phi_i},\quad N_{,ij}={\p^2 N\over \p\phi_i\p\phi_j},\quad N_{,ijk}={\p^3 N\over \p\phi_i\p\phi_j\p\phi_k}.
\q
See \cite{Alabidi:2005qi} in detail. From the Cauchy-Schwarz inequality, the $\tau_{NL}$ is bounded from below by $({6\over 5}f_{NL})^2$,  \cite{Suyama:2007bg},
\e
\tau_{NL}\geq ({6\over 5}f_{NL})^2.
\q
The above inequality is saturated in the single field case, or the vector $N_{,i}$ is an eigenvector of the matrix $N_{,ij}$ in the case with multi fields. So $\tau_{NL}$ is expected to be large if $f_{NL}\gg 1$. Since $N_{,ijk}$ is quite model-dependent, $g_{NL}$ can be negative or positive, and its order of magnitude can be large or small.

The present constraints on the non-Gaussianity parameters from experiments are still loose. For example, WMAP 5yr data \cite{Komatsu:2008hk} implies 
\e
-9<f_{NL}^{local}<111
\q
at $2\sigma$ level. The latest limit on $f_{NL}^{local}$ is 
\e
f_{NL}^{local}=38\pm 21
\q 
at $1\sigma$ level in \cite{Smith:2009jr}. Even though the Gaussian distribution is still consistent with data within $2\sigma$ level, the allowed negative part of $f_{NL}^{local}$ has been cut from the WMAP 3yr data significantly. 

From the theoretical point of view, the curvature perturbations are generated on the superhorizon scale in the curvaton model \cite{Linde:1996gt} and multi-brid inflation model \cite{Sasaki:2008uc,Huang:2009xa} and then a large local-type non-Gaussianity is expected in these two models. As we know, once the direction of inflaton motion changes in the field space during inflation, the entropic fluctuations will make a contribution to the curvature perturbation \cite{Tye:2008ef}. But in this paper we only focus on the curvature perturbation produced at the end of inflation due to the non-trivial condition for inflation to end. For simplicity, we assume that the trajectory of inflaton in the field space is a straight line and then the entropic perturbations do not contribute to the curvature perturbation during inflation.  As long as the path of inflaton during inflation is not orthogonal to the hyper-surface of the end of inflation, the entropic perturbations can contribute to the curvature perturbations at the time when inflation ended. In order to achieve a large non-Gaussianity, this hyper-surface should be curved and the sizes of both bispectrum and trispectrum are determined by how curved this hyper-surface is.

Our paper will be organized as follows. In Sec.~2 we use $\delta N$ formalism to calculate the curvature perturbation in the case with two inflaton fields. In Sec.~3 we investigate the $n$-field inflation model and we find that the calculations of the power spectrum, $f_{NL}$ and $g_{NL}$ are reduced to the two-field case, but $\tau_{NL}$ may encode the information of all of the quantum fluctuations along the $(n-1)$ entropic directions. Some discussions are given in Sec.~4.

\section{Two-field slow-roll inflation}

In this paper, we mainly focus on the curvature perturbation generated at the end of inflation. So we assume that the trajectory of inflaton is a straight line in the field space during inflation. In this section we investigate the two-field slow-roll inflation model with a nontrivial condition for inflation to end. Since the motion of inflaton fields does not change its direction, the equations of motion of the inflatons in the slow-roll inflation takes the form 
\m
{d\phi_i\over dN}= \alpha_i, \quad (i=1,2),
\label{ss}
\n
with constant $\alpha_i$ for $i=1,2$, where $N$ is the number of e-folds before the end of inflation. In this paper we adopt the unit of $M_p=1/\sqrt{8\pi G}=1$. Here we don't need to start with the potential and our results are applicable as long as the equations of motion of inflatons can be written by the above equations.

\begin{figure}[h]
\begin{center}
\includegraphics[width=12cm]{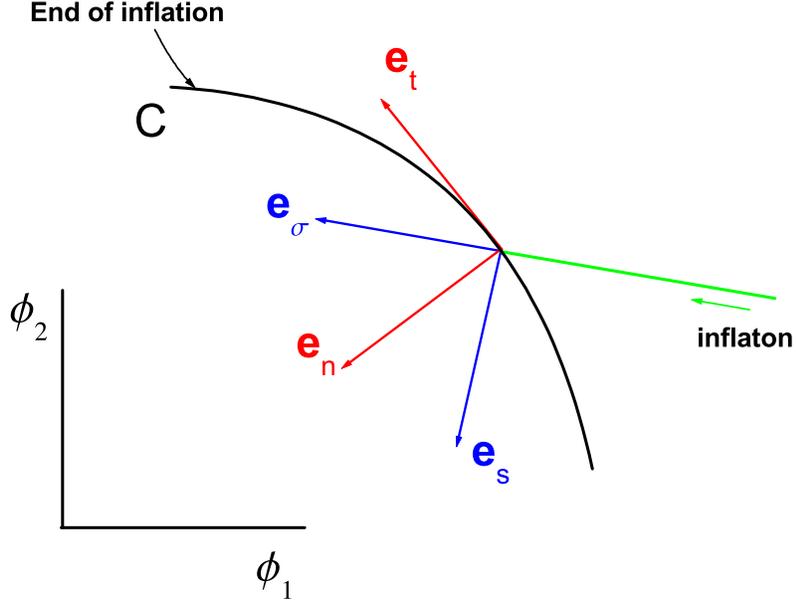}
\end{center}
\caption{The green straight line is the trajectory of inflaton during inflation and the black curve corresponds the field configuration at the end of inflation. }
\label{fig:twof}
\end{figure}

The unit vector along the adiabatic direction (moving direction of inflaton in the field space) is given by
\e
\E_\sigma=-(\alpha_1/\alpha,\alpha_2/\alpha)=(\cos\theta, \sin\theta),
\q
where
\e
\alpha=\sqrt{\alpha_1^2+\alpha_2^2}.
\q
The unit vector along the entropic direction is
\e
\E_s=(-\sin\theta, \cos\theta),
\q
which is orthogonal to $\E_\sigma$. For convenience, the vector of inflaton fields is denoted by
\e
\Phi=(\phi_1,\phi_2),
\q
and its quantum fluctuation is
\e
\delta\Phi=(\delta\phi_1,\delta\phi_2).
\q
The quantum fluctuations along the adiabatic and entropic directions are respectively
\m
\delta\Phi_\sigma&=&\langle \delta\Phi,\E_\sigma \rangle,\\
\delta\Phi_s&=&\langle \delta\Phi,\E_s \rangle.
\n
In this paper, $\langle A, B\rangle$ denotes the inner product of these two vectors $A$ and $B$.
Assume that the condition for the inflation to end is 
\e
F(\phi_{1,f},\phi_{2,f})=0,
\q 
which corresponds to a curve $C$ in the field space. The normal vector of the curve $C$ at $\Phi_f=(\phi_{1,f},\phi_{2,f})$ is given by
\e
{\bar \E}_n=({\p F\over \p \phi_{1,f}}, {\p F\over \p \phi_{2,f}})
\q 
which can be normalized to be 
\e
\E_n={{\bar\E}_n\over ||{\bar\E}_n||},
\q
where
\e
||{\bar\E}_n||=\sqrt{({\p F\over \p \phi_{1,f}})^2+({\p F\over \p \phi_{2,f}})^2}.
\q
We parametrize the curve $C$ to be
\e
\Phi_f=(\phi_{1,f}(s),\phi_{2,f}(s)),
\q
where $s$ is just a parameter. The tangent vector of $C$ at the point $\Phi_f$ is
\e
{\bar\E}_t={d\over ds}\Phi_f=(\phi_{1,f}',\phi_{2,f}'),
\q
which can be normalized to be a unit vector, 
\e
\E_t={{\bar\E}_t \over ||{\bar\E}_t||},
\q
where 
\e
||{\bar\E}_t||=\sqrt{\phi_{1,f}'^2+\phi_{2,f}'^2},
\q
and the prime denotes the derivative with respect to the parameter $s$. For convenience, we introduce a new parameter named arc-length parameter $\ell$ which is related to $s$ by 
\e
d\ell = ||{\bar\E}_t||ds,
\q
and then
\e
\E_t={d\over d\ell}\Phi_f.
\q
In practice it is often difficult to calculate the arc-length parameter, but it is useful for theoretical arguments. The extrinsic curvature of curve associated with $\E_n$ is defined by
\e
\kappa=\langle \E_n, {d\E_t\over d\ell } \rangle.
\q
Since $\E_t$ is a unit vector, the vector ${d\over d\ell}\E_t$ is orthogonal to $\E_t$ and can be taken as a normal vector. According to the above definition of the curvature, we have
\e
{d\over d\ell}\E_t= \kappa \E_n.
\label{dst}
\q
The trajectory of inflaton and the field configuration at the end of inflation in the field space are illustrated in Fig. \ref{fig:twof}.

\begin{figure}[h]
\begin{center}
\includegraphics[width=12cm]{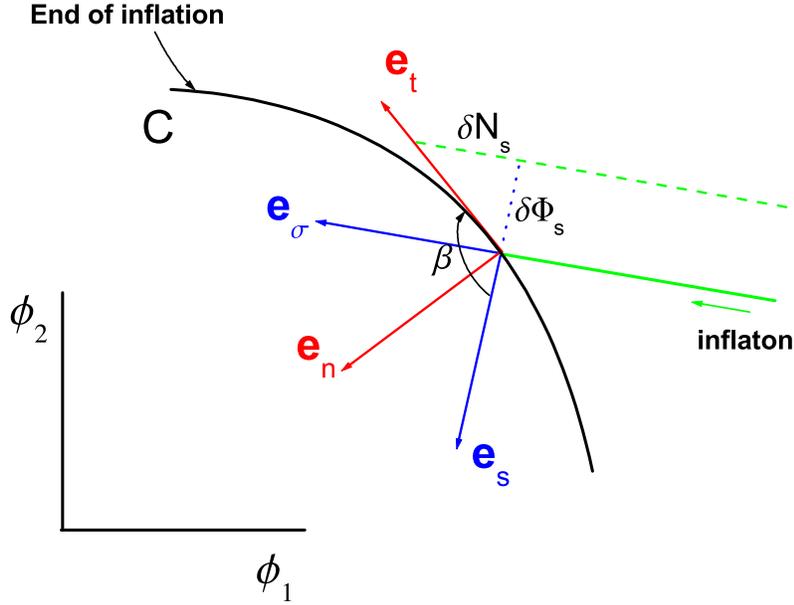}
\end{center}
\caption{The green straight line is the trajectory of inflaton during inflation and the curve corresponds the field configuration at the end of inflation. }
\label{fig:twofdN}
\end{figure}

Using the field vector $\Phi$, the equations of motion can be expressed as
\e
{d\over dN}\Phi=-\alpha \E_\sigma.
\label{mt}
\q
Since $\E_s$ is orthogonal to $\E_\sigma$,  from the equation of motion (\ref{mt}), we have
\e
{d\over dN}\langle \Phi, \E_s \rangle=0.
\q
Thus
\e
\langle \Phi, \E_s \rangle = \langle \Phi_f, \E_s \rangle. 
\label{ppf}
\q 
From Eq.(\ref{ppf}), we have
\e
\delta \ell=\langle D,\delta\Phi\rangle,
\q
where
\e
D=({\p \ell\over \p \phi_1}, {\p \ell\over \p \phi_2})={\E_s\over \langle \E_t, \E_s \rangle}.
\label{funcd}
\q
Integrating over the equation of motion Eq.(\ref{mt}) along the adiabatic direction, we have
\e
\langle \Phi, \E_\sigma \rangle - \langle \Phi_f, \E_\sigma \rangle=-\alpha N,
\q
and then
\e
\langle \delta\Phi, \E_\sigma \rangle - T_1 \langle \delta\Phi, \E_s \rangle=-\alpha \delta N,
\q
where
\e
T_1={\langle \E_t, \E_\sigma \rangle \over \langle \E_t, \E_s \rangle}.
\q
Therefore the curvature perturbation $\delta N$ up to the first order of $\delta \Phi$ is given by
\e
\delta N
%={T_1 \langle \delta\Phi, \E_s \rangle - \langle \delta\Phi, \E_\sigma \rangle \over \alpha} 
= {\langle \delta\Phi, T_1 \E_s-\E_\sigma \rangle \over \alpha}.
\label{stdn}
\q
The minus sign in the term with $\E_\sigma$ comes from the convention in which $N$ denotes the number of e-folds backward in time.
At the leading order, $\delta N$ is contributed from two  sources: one is the adiabatic perturbation $\delta\Phi_\sigma$, the other is the entropic perturbation $\delta\Phi_s$. The reason why 
the entropic perturbation also makes a contribution to the curvature perturbation is that the direction of inflaton movement in the field space is not orthogonal to $C$ due to the nontrivial condition for inflation to end.  Here we give a heuristic explanation on it. See Fig. \ref{fig:twofdN}.
During inflation the inflatons evolve along the solid green line classically. Once we take the quantum fluctuations into account, the trajectory, illustrated by the dashed green line, slightly deviates from the classical one.  If the entropic direction $\E_s$ does not parallel to the tangent direction $\E_t$, an additional contribution to the $\delta N$ appears,  
\e
\delta N_s=\tan\beta{\delta\Phi_s\over \alpha},
\q
where $\tan \beta$ is nothing but $T_1$. If the path of inflaton is orthogonal to $C$, $\langle \E_\sigma, \E_t \rangle=T_1=0$ and then the entropic perturbation does not contribute to the curvature perturbation. If so, the two-field inflation is reduced to the single-field case.

From Eq.(\ref{stdn}), we get
\e
N_{,i}={T_1\E_s^i-\E_\sigma^i \over \alpha}.
\q
Here we need to stress that $T_1$ contains parameteter $\ell$ which is related to $\Phi$ through $\E_t$. In order to calculate the bispectrum and trispectrum, we need to expand the number of e-folds to higher orders. From the above equation, we get
\e
N_{,ij}={1\over \alpha}{\p T_1\over \p \ell}\E_s^i D_j.
\q
Taking Eq.(\ref{dst}) into account, we have
\e
{\p T_1\over \p \ell}=\kappa T_2, 
\q
where $\kappa$ is the extrinsic curvature of curve $C$ at $\Phi_f$,
\e
T_2=\hbox{sign}{1 \over \langle \E_t, \E_s\rangle^2 },
\q
and 
\e
\hbox{sign}=\langle \E_n, \E_\sigma\rangle \langle \E_t, \E_s \rangle - \langle \E_t, \E_\sigma\rangle \langle \E_n, \E_s \rangle.
\q
Here $\hbox{sign} =\pm$: `+' corresponds to the case with $\E_t\times \E_n =-\E_\sigma \times \E_s$, and `-'  corresponds to the case with $\E_t\times \E_n =\E_\sigma \times \E_s$. Here $A\times B=\epsilon_{ij}A^iB^j$ and $\epsilon_{12}=-\epsilon_{21}=1$. Therefore 
\e
N_{,ij}=\hbox{sign}{\kappa\over \alpha} {\E_s^i \E_s^j \over \langle \E_t, \E_s\rangle^3}.
\q
Here both $\kappa$ and $\E_t$ are the functions of $\Phi$. Similarly, we have
\e
N_{,ijk}=-\hbox{sign}{\kappa^2 \over \alpha} \(\tau+3 {\langle \E_n, \E_s \rangle \over \langle \E_t, \E_s\rangle}\){\E_s^i \E_s^j \E_s^k \over \langle \E_t, \E_s\rangle^4},
\q 
where
\e
\tau={d\kappa^{-1} \over d\ell},
\q
which measures the variation of the curvature radius per unit arc-length.
For a curve with a constant curvature, $\tau=0$. Here we need to stress that all of the non-linear orders of the curvature perturbation $\delta N$ come from the entropic perturbations which are converted into the curvature perturbation at the end of inflation. Therefore the distribution of curvature perturbation becomes Gaussian if the path of inflatons is orthogonal to the curve $C$ of the end of inflation in our scenario.

Assume that the scalar field fluctuations $\delta\phi_1$ and $\delta\phi_2$ are Gaussian and non-correlated, 
\e
\langle \delta\phi_i \delta\phi_j \rangle=\({H_*\over 2\pi}\)^2\delta_{ij},
\q
where $H_*$ denotes the Hubble parameter at the time of Hubble exit during inflation. The amplitude of the primordial curvature perturbation is
\e
P_\zeta={1\over \alpha^2\langle \E_t, \E_s\rangle^2}({H_*\over 2\pi})^2.
\q
Here we consider that 
\e
\langle \E_t, \E_\sigma \rangle^2+\langle \E_t, \E_s\rangle^2=1.
\label{etss}
\q
The amplitude of the gravitational wave fluctuations is related to the inflation scale by
\e
P_T=8({H_*\over 2\pi})^2,
\q
and then the tensor-scalar ratio takes the form
\e
r=P_T/P_\zeta=8\alpha^2 \langle \E_t, \E_s\rangle^2.
\q
The equations of motion in Eq.(\ref{ss}) can be rewritten by
\e
{d\phi_i\over dN}={1\over V}{\p V\over \p \phi_i}={\p \ln H^2 \over \p \phi_i}=\alpha_i,
\q
and then the spectral index of the scalar power spectrum is 
\e
n_s=1+{d\ln P_\zeta \over d\ln k}= 1-{d\ln P_\zeta \over dN}=1-\alpha^2.
\q
A red-tilted primordial power spectrum is predicted. 
Since $\langle \E_t, \E_s\rangle^2\leq 1$, 
\e
r\leq 8(1-n_s).
\q
The tensor-scalar ratio in this model is bounded from above by the deviation from the exact scalar invariance. Similarly, the spectral index of the tensor perturbation is given by
\e
n_T=-\alpha^2=-{r\over 8 \langle \E_t, \E_s\rangle^2},
\q
which is bounded from above by the tensor-scalar ratio,
\e
n_T\leq -{r\over 8}.
\q
In single-field inflation model, the consistency relation is $n_T=-r/8$ which saturates the upper bound in the above inequality. In the multi-field case, the spectrum of the tensor perturbation can be more red-tilted.

From Eq.(\ref{mt}), 
\e
{\langle \Delta \Phi, \E_\sigma\rangle }=-\alpha \Delta N
\q
which measures the distance of inflaton excursion along the adiabatic direction in the field space in unit of $M_p$. Now the the tensor-scalar ratio becomes 
\e
r=8\langle \E_t, \E_s\rangle^2\({\langle \Delta \Phi, \E_\sigma\rangle \over \Delta N}\)^2,
\q
which is smaller than that in the single-field model \cite{Lyth:1996im} when $\langle \E_t, \E_s\rangle^2< 1$. The reason is that the entropic perturbation also makes a contribution to the total curvature perturbation and then the tensor-scalar ratio becomes smaller compared to the case with only the adiabatic perturbation.

\begin{figure}[h]
\begin{center}
\includegraphics[width=12cm]{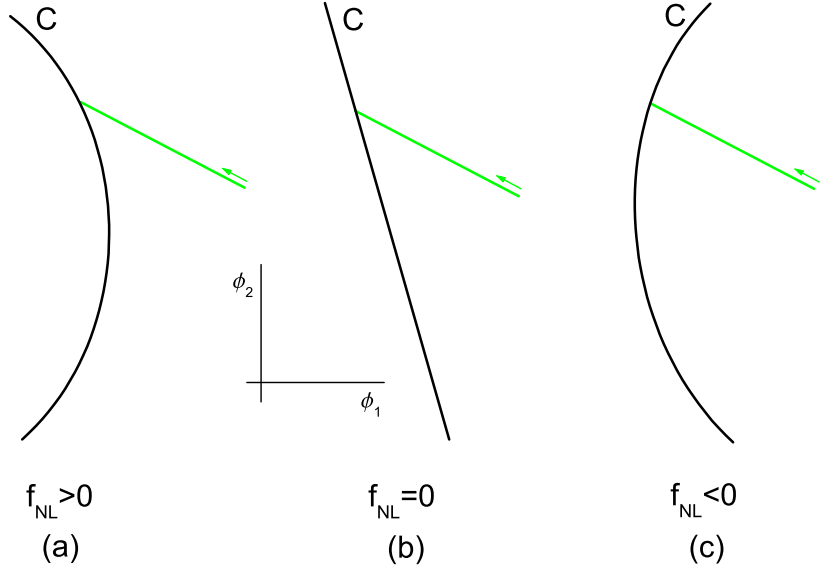}
\end{center}
\caption{The green straight line is the trajectory of inflaton fields during inflation and the curve $C$ corresponds the field configuration at the end of inflation. $C$ is a straight line in (b).}
\label{fig:twofnl}
\end{figure}

We can also easily calculate the non-Gaussianity parameters,
\m
f_{NL}&=& \hbox{sign}{5\over 6} \cdot\alpha \kappa {\langle \E_t, \E_\sigma \rangle^2 \over \langle \E_t, \E_s\rangle},\\
\tau_{NL}&=&\alpha^2 \kappa^2  {\langle \E_t, \E_\sigma \rangle^2 \over \langle \E_t, \E_s\rangle^2 },\\
g_{NL}&=&-\hbox{sign}{25 \over 54} \cdot \alpha^2\kappa^2\(\tau+3 {\langle \E_n, \E_s \rangle \over \langle \E_t, \E_s\rangle}\) {\langle \E_t, \E_\sigma \rangle^3 \over \langle \E_t, \E_s \rangle}.
\n
If the curve $C$ corresponding to the inflaton field configuration at the end of inflation is a straight line, $\kappa=0$ and hence all of the non-Gaussianity parameters are equal to zero. Here we need to point out that $\kappa$ is not definitely positive and $f_{NL}$ can be positive or negative. The sign of $f_{NL}$ is illustrated in Fig.\ref{fig:twofnl}.
In the case (b), even though the entropic perturbation contributes to the total curvature perturbation at the end of inflation if the path of inflaton is not orthogonal to the curve $C$,  this contribution linearly depends on the quantum fluctuations of inflation fields and then the curvature perturbations at all of the non-linear orders are still equal to zero. Keeping $f_{NL}$ fixed, we have
\m
\tau_{NL}&=&{1\over \langle \E_t, \E_\sigma \rangle^2}({6\over 5}f_{NL})^2,\\
g_{NL}&=&\(2-\hbox{sign}{2\over 3}\cdot \tau{\langle \E_t, \E_s \rangle \over \langle \E_t, \E_\sigma \rangle }\)f_{NL}^2.
\n
Since $\langle \E_t, \E_\sigma \rangle^2 \leq 1$, 
\e
\tau_{NL}\geq ({6\over 5}f_{NL})^2,
\q
which is consistent with the result in \cite{Suyama:2007bg}. If $C$ is a curve with constant curvature, $\tau=0$ and then $g_{NL}=2f_{NL}^2$.

From Eq.(\ref{etss}), we have
\e
f_{NL}=\hbox{sign}{5\over 6} \cdot\alpha \kappa \({1\over \langle \E_t, \E_s\rangle} - \langle \E_t, \E_s\rangle \).
\q
WMAP 5yr data \cite{Komatsu:2008hk} implies 
\e
n_s=0.96_{-0.013}^{+0.014}.
\q
Considering that $\langle \E_t, \E_s\rangle^2$ is related to $r$ and $n_s$ by
\e
\langle \E_t, \E_s\rangle^2={r\over 8(1-n_s)},
\q
for $n_s=0.96$, if $r=0.32 \langle \E_t, \E_s\rangle^2\ll1$, we have
\e
f_{NL}\simeq \hbox{sign}{5\over 6} \cdot {\alpha \kappa \over \langle \E_t, \E_s\rangle}.
\q
If $r\ll1$, $\langle \E_t, \E_\sigma \rangle^2\simeq 1$ and we have
\e
\tau_{NL}\simeq ({6\over 5}f_{NL})^2.
\q
We see that $\tau_{NL}$ is roughly equal to $({6\over 5}f_{NL})^2$ for the low-scale inflation model.
However, $\tau={d \kappa^{-1} / d\ell}$ is a model-dependent parameter and $g_{NL}$ can be positive or negative and its order of magnitude can be large or small.  

As a check, we use the above formula to calculate the curvature perturbation in the model in \cite{Huang:2009xa} and we find that our results in this paper are the same as those in \cite{Huang:2009xa}.

\section{$n$-field slow-roll inflation}

In this section, we consider the inflation model with $n$ inflaton fields. Because we only focus on the curvature perturbation generated at the end of inflation, we still assume that the trajectory of inflaton in the field space is a straight line during inflation. The equations of motion of inflatons are given by
\m
{d\phi_i\over dN}= \alpha_i, 
\n
with constant $\alpha_i$ for $i=1,2,...,n$. The unit vector along the adiabatic direction is 
\e
\E_\sigma=-(\alpha_1,\alpha_2,...,\alpha_n)/\alpha,
\q
where 
\e
\alpha=\(\sum_{i=1}^n \alpha_i^2\)^\half.
\q

\begin{figure}[h]
\begin{center}
\includegraphics[width=12cm]{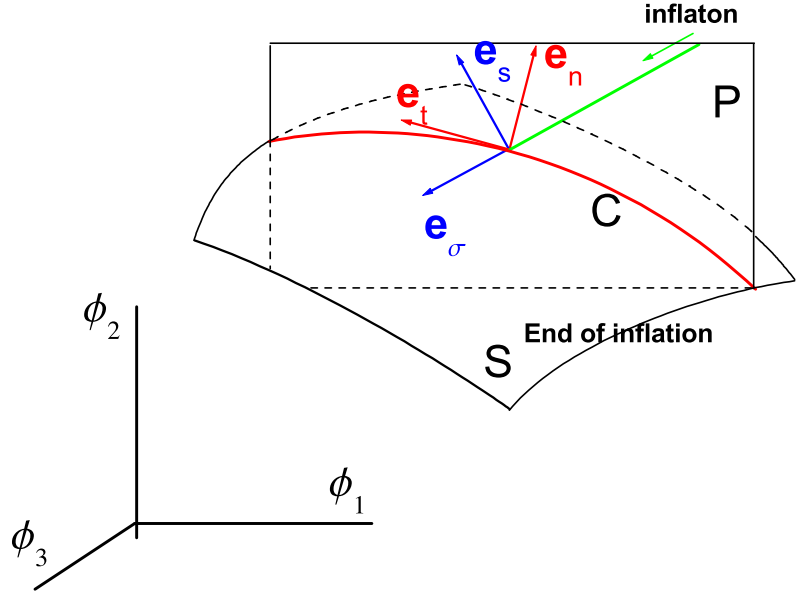}
\end{center}
\caption{The green line is the trajectory of inflaton fields during inflation and the hyper-surface corresponds the field configuration at the end of inflation. }
\label{fig:fields}
\end{figure}

Similarly, we also denote the inflaton field vector in the field space as
\e
\Phi=(\phi_1,\phi_2,...,\phi_n),
\q
and its quantum fluctuation is 
\e
\delta\Phi=(\delta\phi_1,\delta\phi_2,...,\delta\phi_n).
\q
The equation of motion can be written by
\e
{d\Phi\over dN}=-\alpha\E_\sigma.
\label{npns}
\q
Assume that the values of inflaton fields at the end of inflation are given by
\e
\Phi_f=(\phi_{1,f}, \phi_{2,f},..., \phi_{n,f}),
\q
which satisfies the equation
\e
F(\Phi_f)=F(\phi_{1,f}, \phi_{2,f},..., \phi_{n,f})=0. 
\q
The solution of the above equation is described by a $(n-1)$-dimensional hyper-surface $S$ in the $n$-dimensional field space. 
This hyper-surface $S$ associated with the field configuration at the end of inflation in the field space has a normal vector at $\Phi_f$ as
\e
{\bar\E}_n=({\p F\over \p \phi_{1,f}}, {\p F\over \p \phi_{2,f}}, ... , {\p F\over \p \phi_{n,f}}),
\q
which can be normalized to be
\e
\E_n={{\bar\E}_n \over ||{\bar\E}_n||},
\q
where
\e
||{\bar\E}_n||=\(\sum_{i=1}^n ({\p F\over \p \phi_{i,f}})^2\)^\half.
\q
There are $(n-1)$ independent vectors along the entropic directions which are orthogonal to the adiabatic direction. In particular, one of them is very important, namely $\E_s$ who stays on the plane $P$ determined by $\E_n$ and $\E_\sigma$. Assume $\E_\sigma$ is not orthogonal to the hyper-surface $S$; otherwise, the entropic perturbations cannot contribute to the curvature perturbation and this $n$-field inflation model is reduced to the single-field case. Now the vector along this special entropic direction can be given by
\e
{\bar\E}_s=\E_n-\langle \E_n, \E_\sigma \rangle \E_\sigma
\q 
which can be normalized to be a unit vector
\e
\E_s={{\bar \E}_s \over ||{\bar \E}_s||},
\q
with
\e
||{\bar\E}_s||=\sqrt{1-\langle \E_\sigma, \E_n\rangle^2}.
\q
Similarly, the unit tangent vector of $S$ on the plane $P$ can be expressed by
\e
\E_t={{\bar \E}_t \over ||{\bar \E}_t||},
\q
where
\e
{\bar\E}_t=\E_\sigma-\langle  \E_n, \E_\sigma \rangle \E_n,
\q
and 
\e
||{\bar \E}_t||=\sqrt{1-\langle \E_\sigma, \E_n\rangle^2}.
\q
Here the vectors $\E_t$ and $\E_s$ are constructed by $\E_\sigma$ and $\E_n$, and then we obtain two simple relations,
\e
\langle \E_t, \E_s \rangle = -\langle \E_n, \E_\sigma \rangle,
\label{tsns}
\q
and
\e
\langle \E_n, \E_\sigma\rangle \langle \E_t, \E_s \rangle - \langle \E_t, \E_\sigma\rangle \langle \E_n, \E_s \rangle=-1.
\q
These vectors are illustrated in Fig. \ref{fig:fields}.
The curve $C$ is the intersection curve between the hyper-surface $S$ and the plane $P$. 

There are still $(n-2)$ independent entropic directions orthogonal to the plane $P$. The unit vectors along these $(n-2)$ entropic directions are denoted by ${\E}_{s,A}$ for $A=1,2,..,n-2$. Since ${\E}_{s,A}$ is orthogonal to $\E_\sigma$ and $\E_s$ which are the two independent vectors living on the plane $P$, ${\E}_{s,A}$ must be orthogonal to the plane $P$ and then ${\E}_{s,A}$ for $A=1,2,...,n-2$ are also the unit tangent vectors of $S$ at the point $\Phi_f$, namely 
\e
{\E}_{t,A}={\E}_{s,A}.
\q 
Denote $\ell$ as the arc-length parameter along the tangent direction $\E_t$ and ${\ell}_A$ as the arc-length parameter along the tangent direction ${\E}_{t,A}$ for $A=1,2,...,n-2$. Therefore
\e
\E_t={\p \Phi_f \over \p\ell}, \quad {\E}_{t,A}={\p \Phi_f \over \p{\ell}_A}.
\q

Keeping ${\E}_{s,A}$ fixed and integrating over Eq.(\ref{npns}) along the these $(n-2)$ entropic directions, we obtain
\e
\langle \Phi,\E_{s,A}\rangle=\langle \Phi_f,\E_{s,A}\rangle,
\q
and then
\m
\delta {\ell}_A&=&\langle {{\E}_{s,A}\over \langle {\E}_{t,A}, \E_{s,A}\rangle}, \delta\Phi \rangle=\langle {\E}_{s,A}, \delta\Phi \rangle,
\n
at the linear level. However, we need to expand $\delta \ell$ up to the first non-linear order,
\m
\langle \delta\Phi, \E_s\rangle&=&\langle \delta\Phi_f, \E_s\rangle \nonumber \\
&=&\langle \E_t,\E_s \rangle \delta\ell+\half \kappa \langle \E_n,\E_s\rangle (\delta \ell)^2+\half \sum_{A=1}^{n-2} (2\kappa_A\langle \E_n,\E_s\rangle + \vartheta_{A}\langle \E_t,\E_s\rangle)\delta\ell\delta\ell_A,
\label{sell}
\n
where 
\m
\kappa&=&\langle {\p\E_t\over \p\ell},\E_n\rangle,\\
\vartheta_{A}&=&\langle {\p\E_{t,A}\over \p\ell},\E_t\rangle,
\n
and 
\e
\kappa_A=\langle {\p\E_t\over \p\ell_A},\E_n\rangle = \langle {\p\E_{t,A}\over \p\ell},\E_n\rangle
\q
is the offdiagonal component of the extrinsic curvature tensor of the hyper-surface $S$ along $\E_t$ and $\E_{t,A}$ at the point $\Phi_f$. 
Eq.(\ref{sell}) can be perturbatively solved as
\m
\delta \ell&=&{1 \over \langle \E_t,\E_s \rangle}\langle \delta\Phi, \E_s\rangle  
-{\kappa\over 2}{\langle \E_n,\E_s \rangle \over \langle \E_t,\E_s\rangle^3}\langle \delta\Phi, \E_s\rangle^2 \nonumber \\ 
&-&{1\over 2\langle \E_t,\E_s\rangle^2}\sum_{A=1}^{n-2}  (2\kappa_A\langle \E_n,\E_s\rangle + \vartheta_{A}\langle \E_t,\E_s\rangle) \langle \delta\Phi, \E_s\rangle\langle \delta\Phi, \E_{s,A}\rangle.
\n
Integrating over Eq.(\ref{npns}) along the adiabatic direction, we have 
\e
\langle \Phi, \E_\sigma \rangle - \langle \Phi_f, \E_\sigma \rangle=-\alpha N,
\label{efold}
\q
and $\delta N$ is expanded up to the first non-linear order as
\m
\delta N&=& - {1\over \alpha}\langle \delta\Phi,\E_\sigma\rangle+{1\over \alpha}\langle \E_t,\E_\sigma\rangle \delta \ell  \nonumber \\
&+&{\kappa \over 2\alpha} \langle \E_n,\E_\sigma\rangle (\delta \ell)^2+{1\over 2\alpha}\sum_{A=1}^{n-2} (2\kappa_A\langle \E_n,\E_\sigma \rangle + \vartheta_{A}\langle \E_t,\E_\sigma \rangle)\delta\ell\delta\ell_A 
+{\cal O}(\delta \ell_A\delta \ell_B) \nonumber \\
&=&{1\over \alpha}\langle T_1\E_s-\E_\sigma, \delta\Phi \rangle \nonumber \\
&-&{\kappa \over 2\alpha} {1\over \langle \E_t,\E_s\rangle^3} \langle \delta\Phi,\E_s \rangle^2 - {1\over \alpha}{1\over \langle \E_t,\E_s\rangle^2} \sum_{A=1}^{n-2}\kappa_A \langle \delta \Phi,\E_{s,A}\rangle \langle \delta \Phi,\E_s\rangle \nonumber \\
&+&{\cal O}(\langle \delta \Phi,\E_{s,A}\rangle \langle \delta \Phi,\E_{s,B}\rangle)
\label{dtnl}
\n
where
\m
T_1={\langle \E_t,\E_\sigma\rangle \over \langle \E_t,\E_s \rangle}.
\n

At the linear level, we obtain
\e
(\delta N)_L={1\over \alpha}\langle T_1\E_s-\E_\sigma, \delta\Phi \rangle.
\q
Therefore, at the linear level, the calculation of the curvature perturbation generated at the end of $n$-field inflation has been simplified to be that for two-field inflation model, and 
\e
N_{,i}={T_1\E_s^i-\E_\sigma^i\over \alpha}.
\q
However, the perturbation along the $(n-2)$ entropic directions ${\E}_{s,A}$ which is orthogonal to the plan $P$ appear in the non-linear order of the expansion of the number of e-folds $N$. From Eq.(\ref{nfnl}) and (\ref{ngnl}), both $f_{NL}$ and $g_{NL}$ are proportional to $N_{,i}$ and hence all of the perturbations along ${\E}_{s,A}$ do not contribute to $f_{NL}$ and $g_{NL}$ because $\langle \E_s, \E_{s,A}\rangle=\langle \E_\sigma, \E_{s,A}\rangle=0$ for $A=1,2,...,n-2$. But the perturbations along these $(n-2)$ entropic directions may make some contributions to $\tau_{NL}$. We need to calculate $\tau_{NL}$ carefully.
From Eq.(\ref{dtnl}),
\m
N_{,ij}=-{\kappa \over \alpha}{\E_s^i\E_s^j\over \langle \E_t,\E_s\rangle^3}-{1\over \alpha}{1\over \langle \E_t,\E_s\rangle^2}\sum_{A=1}^{n-2} \kappa_A (\E_s^i\E_{s,A}^j+\E_s^j\E_{s,A}^i)+{\cal O}(\E_{s,A}^i\E_{s,B}^j).
\n
We can also easily obtain
\e
N_{,ijk}={\kappa^2\over \alpha}\(\tau+3{\langle \E_n,\E_s\rangle \over \langle \E_t, \E_s\rangle}\){\E_s^i\E_s^j\E_s^k\over \langle \E_t,\E_s\rangle^4}+{\cal O}(\E_{s,A}^i).
\q
Here we need to stress that ${\cal O}(\E_{s,A}^i\E_{s,B}^j)$ in $N_{,ij}$ and ${\cal O}(\E_{s,A}^i)$ in $N_{,ijk}$ are complicated, and do not make any contribution to the bispectrum and trispectrum. So we don't work them out.

Taking Eq.(\ref{tsns}) into account, we find that the amplitude of the primordial power spectrum can be re-written by
\e
P_\zeta={1\over \alpha^2\langle \E_n, \E_\sigma\rangle^2}({H_*\over 2\pi})^2,
\q
and the non-Gaussianity parameters become
\m
f_{NL}&=& {5\over 6} \cdot\alpha \kappa {\langle \E_t, \E_\sigma \rangle^2 \over \langle \E_n, \E_\sigma \rangle},\\
\tau_{NL}&=&\alpha^2 \(\kappa^2 + \kappa_s^2\langle \E_n,\E_\sigma\rangle^2 \){\langle \E_t, \E_\sigma \rangle^2 \over \langle \E_n, \E_\sigma\rangle^2 },\\
g_{NL}&=&-{25 \over 54} \cdot \alpha^2\kappa^2\(\tau-3 {\langle \E_n, \E_s \rangle \over \langle \E_n, \E_\sigma \rangle}\) {\langle \E_t, \E_\sigma \rangle^3 \over \langle \E_n, \E_\sigma \rangle},
\n
where
\e
\kappa_s^2=\sum_{A=1}^{n-2}\kappa_A^2.
\q
The curvature perturbation up to the third order has been computed completely for this $n$-field inflation model. Since $\kappa_s^2\langle \E_n,\E_\sigma \rangle^2\geq 0$, $\tau_{NL}\geq ({6\over 5}f_{NL})^2$. One can also apply this constructive method to calcualte the curvature perturbation in the two-fields inflation in Sec.~2.

An interesting observation is that if $\kappa=0$ and the inflaton path is not orthogonal to the hyper-surface $S$ both $f_{NL}$ and $g_{NL}$ are equal to zero, but $\tau_{NL}$ can be non-zero. This is different from that in the two-field case and the curvaton model \cite{Linde:1996gt}. In this special case, we find
\e
\tau_{NL}=(1-n_s-{r\over 8}) \kappa_s^2.
\q
If $r\ll 1$, $\tau_{NL}\simeq (1-n_s)\kappa_s^2$.

Usually it is not easy to parametrize the curve $C$ in the $n$-dimensional field space. Here we will provide a general method to calculate the relevant geometric quantities associated with the curve $C$ and the hyper-surface $S$. Assume that the orthogonal coordinates on the  hyper-surface $S$ are denoted by
\e
s=(s_1,s_2,...,s_{n-1}),
\q 
and then $\Phi_f$ is a function of $s_1,s_2,...,s_{n-1}$. There are $(n-1)$ tangent vectors of $S$ at point $\Phi_f$,
\e
{\bar\E}_{t,p}={\p \Phi_f\over \p s_q}=({\p \phi_{1,f}\over \p s_q}, {\p \phi_{2,f}\over \p s_q}, ... , {\p \phi_{n,f}\over \p s_q}),
\q
for $q=1,2,...,n-1$. These vectors can be normalized to be the unit vectors,
\e
\E_{t,q}={{\bar\E}_{t,q} \over ||{\bar\E}_{t,q}||},
\q
where
\e
||{\bar\E}_{t,q}||=\(\sum_{i=1}^n ({\p \phi_{i,f}\over \p s_q})^2 \)^\half.
\q
For convenience, we introduce $(n-1)$ arc-length parameters $\ell_q$ which satisfy
\e
d\ell_q=||{\bar\E}_{t,q}||d s_q,
\q
and then
\e
\E_{t,q}={\p\Phi_f \over \p\ell_q}.
\q
Since the set of $\E_{t,q}$ for $q=1,2,...,n-1$ is an orthogonal and complete frame in the tangent space of $S$, $\E_t$ can be expanded by
\e
\E_t=\sum_{q=1}^{n-1}\langle \E_t,\E_{t,q}\rangle \E_{t,q}.
\label{etq}
\q
Since $\E_t={\p \Phi_f \over \p\ell}$, from Eq.(\ref{etq}), we have
\e
{\p\over \p\ell}=\sum_{q=1}^{n-1}\langle \E_t,\E_{t,q}\rangle {\p \over \p\ell_q}.
\label{dlq}
\q
The relevant geometric quantities associated with the curve $C$, such as the curvature $\kappa$ and $\tau$, can be calculated by using the formulation in Eq.(\ref{dlq}). For example,
\m
\kappa=\langle \E_n, \sum_{q=1}^{n-1} \langle \E_t,\E_{t,q}\rangle {\p \E_t\over \p\ell_q} \rangle.
\n
The value of $\kappa_s^2$ can be easily calculated as follows,
\e
\kappa_s^2=\kappa_T^2-\kappa^2,
\q
where 
\e
\kappa_T^2=\sum_{p=1}^{n-1} \langle {\p\E_{t,p}\over \p\ell},\E_n \rangle^2.
\q
Now, in principle, we work out all of the non-Gaussianity parameters for the n-field inflation.

%On the other hand, if $\E_\sigma$ parallels to $\E_n$, all of the non-linear orders of the curvature %perturbation are zero and hence all of the non-Gaussianity parameters are equal to zero. 

%\e
%\kappa=n_p t^q \p_q t^p.
%\q

\section{Discussions}

The three-point correlation function of curvature perturbations has been a sensitive probe of the physics in the early universe. In the near future even the four-point correlation function can be measured if it is not so small. Now the investigation of the primordial curvature perturbation up to at least the second non-linear order becomes more and more important. Recently many relevant papers concerning the primordial non-Gaussianity also appear in \cite{Huang:2008ze}. If a large non-Gaussianity is confirmed by the upcoming experiments, the single-field slow-roll inflation will be ruled out.

In the single-field slow-roll inflation model, usually the inflation is assumed to end when the slow-roll condition is violated. However the inflation terminated by a water-fall field is more generic in the multi-field inflation model. In particular, it is quite natural that a tachyonic field appears towards the end of inflationary epoch in many scenarios inspired by string theory. In this paper, for simplicity, the trajectory of inflaton is assumed to be a straight line in the field space during inflation. The quantum fluctuations along the entropic directions can contribute to the total curvature perturbation as long as the trajectory of inflaton is not orthogonal to the hyper-surface of the end of inflation. The size of the non-Gaussianity is related to how curved this hyper-surface is. In this scenario, the primordial bispectrum and trispectrum have a local shape because both of them are generated on the super-horizon scale. If the inflaton path is not a straight line, it is much more difficult to calculate the curvature perturbation analytically. In addition, it is also possible that the trajectory in field space has not yet converged at the end of inflation and then the curvature perturbation still evolves during reheating. We will come back to these cases in the future.

The local-shape bispectrum and trispectrum are characterized by three non-Gaussianity parameters: $f_{NL}$, $\tau_{NL}$ and $g_{NL}$. In principle, they are three independent parameters. Here we want to stress that both $f_{NL}$ and $g_{NL}$ can be positive or negative respectively, but $\tau_{NL}$ must be positive and not less than $({6\over 5}f_{NL})^2$. If there are three or more inflaton fields, $\tau_{NL}$ can be positively large even when $f_{NL}=g_{NL}=0$. This is quite different from the curvaton model.

Here we also want to give a simple realization of our model. For example, the inflation potential takes the form
\e
V(\Phi)\equiv V(\phi_1,...,\phi_n)=V_0\exp\(\sum_{i=1}^n\alpha_i\phi_i \),
\q
where
\e
V_0={\lambda\over 4}\(\chi^2-{m^2\over \lambda}\)^2+\half [F(\Phi)+m^2]\chi^2.
\q
During inflation $F(\Phi)$ is assumed to be positive and the field $\chi$ is trapped at $\chi=0$. When $F(\Phi)=0$ the field $\chi$ becomes a waterfall field and the inflation ended. We can easily check that the slow-roll equation of motion of inflaton field $\Phi$ is given by Eq.(\ref{npns}). The slow-roll condition is nothing but $\alpha_i^2\ll 1$. Usually the hyper-surface of $F(\Phi)=0$ is not a uniform density slice and the hot state of the universe starts at slightly different temperatures for different values of $\Phi_f$ at which inflation ended. A correction to the total number of e-folds is needed, namely
\e
N_c={1\over 4}\ln \[{V(\Phi_f)\over V_0}\]=-{\alpha\over 4} \langle \E_\sigma, \Phi_f \rangle.
\q
Combining with Eq.(\ref{efold}), we obtain
\e
N=-{1\over \alpha}\langle \Phi,\E_\sigma\rangle+\({1\over \alpha}-{\alpha\over 4}\)\langle \Phi_f, \E_\sigma \rangle.
\q
Since $\alpha\ll 1$, the correction to the total number of e-folds $N_c$ can be ignored. Approximately, we calculate the non-linear parameters defined for the curvature perturbation on the uniform density slice.

In this paper we find a correspondence between the non-Gaussianity parameters and the geometric quantities of the hyper-surface for inflation to end. The non-Gaussianity parameters can be measured by the experiments and then we can easily re-construct the geometry of this hyper-surface in the field space. A more ambitious project is to build up a connection between the geometry of this hyper-surface and the geometry of the compactified space in string theory. If this project can be achieved, it seems that we can directly ``measure" the compactified space. Here we also need to point out one shortcoming: the non-Gaussianity parameters only depend on the local geometry around the point where the inflation ends in the field space and the information about the geometry we can get is still limited. Anyway, we hope that the non-Gaussianity will provide more information for the fundamental physics in the near future.

\vspace{1.4cm}

\noindent {\bf Acknowledgments}

\vspace{.5cm}

We would like to thank Xian Gao, Bin Hu, Miao Li, Yushu Song, Erick Weinberg and Piljin Yi for useful discussions.

%\vspace{1.5cm}

%\appendix

%\section{Basic knowledge about a line in a two-dimensional geometry}
%\label{ap}

\end{document}